\documentclass{scrartcl}

\usepackage{amssymb}
\usepackage{amsthm}
\usepackage{amsmath}
\usepackage{hyperref}
\usepackage{doi}

\newcommand{\Ri}[1]{\mathfrak{R}(#1)}
\newcommand{\Li}[1]{\mathfrak{L}(#1)}
\newcommand{\RInv}[1]{\overrightarrow{\mathfrak{I}}(#1)}
\newcommand{\LInv}[1]{\overleftarrow{\mathfrak{I}}(#1)}
\newcommand{\E}[1]{\mathfrak{E}(#1)}
\newcommand{\rev}[1]{{#1}^{\mathrm{rev}}}
\newcommand{\opp}[1]{{#1}^{\mathrm{op}}}

\newcommand{\T}{\mathcal{T}}

\newcommand{\defeq}{:=}

\newcommand{\One}{\mathbf{1}}
\newcommand{\step}[1]{\Rightarrow_{#1}}

\newcommand{\N}{\mathbb{N}}

\newcommand{\Q}{\mathbb{Q}}

\newcommand{\factors}[2]{\rho(#2, #1)}

\newtheorem{theorem}{Theorem}
\newtheorem{lemma}[theorem]{Lemma}

\def\shuf{\mathbin{\mathchoice
{\rule{.03em}{1ex}\rule{.3em}{.03em}\rule{.03em}{1ex}
 \rule{.3em}{.03em}\rule{.03em}{1ex}}%
{\rule{.03em}{1ex}\rule{.3em}{.03em}\rule{.03em}{1ex}
 \rule{.3em}{.03em}\rule{.03em}{1ex}}%
{\rule{.02em}{.7ex}\rule{.2em}{.02em}\rule{.2pt}{.7ex}
 \rule{.2em}{.02em}\rule{.02em}{.7ex}}%
{\rule{.03em}{1ex}\rule{.3em}{.03em}\rule{.03em}{1ex}
 \rule{.3em}{.03em}\rule{.03em}{1ex}}%
}}

\begin{document}

\title{On the capabilities of grammars, automata, and transducers controlled by monoids}
\author{Georg Zetzsche}
\titlehead{Fachbereich Informatik, Technische Universit\"{a}t Kaiserslautern, \\
Postfach 3049, 67653 Kaiserslautern, Germany \\
\texttt{zetzsche@cs.uni-kl.de}}

\maketitle

\begin{abstract}
During the last decades, classical models in language theory have been extended
by control mechanisms defined by monoids.  We study which monoids cause the
extensions of context-free grammars, finite automata, or finite state
transducers to exceed the capacity of the original model.  Furthermore, we
investigate when, in the extended automata model, the nondeterministic variant
differs from the deterministic one in capacity.  We show that all these
conditions are in fact equivalent and present an algebraic characterization. In
particular, the open question of whether every language generated by a valence
grammar over a finite monoid is context-free is provided with a positive
answer. 
\end{abstract}

\section{Introduction}

The idea to equip classical models of theoretical computer science with a
monoid (or a group) as a control mechanism has been pursued by several authors
in the last decades
\cite{FernauStiebe2002a,ItoMartinVideMitrana2001,Kambites2009,MitranaStiebe2001,Paun1980,RenderKambites2009}.
This interest is justified by the fact that these extensions allow for
a uniform treatment of a wide range of automata and grammar models: Suppose a
storage mechanism can be regarded as a set of states on which a set of
partial transformations operates and a computation is considered valid
if the composition of the executed transformations is the identity.
Then, this storage constitutes a certain monoid control. 

For example,
in a pushdown storage, the operations \emph{push} and \emph{pop} (for
each participating stack symbol) and
compositions thereof are partial transformations on the set of words
over some alphabet.  In this case, a computation is considered valid
if, in the end, the stack is brought back to the initial state, i.e.,
the identity transformation has been applied.
As further examples, blind and partially blind
multicounter automata (see \cite{Greibach1978}) can be regarded as finite automata controlled by a
power of the integers and of the bicyclic monoid (see
\cite{RenderKambites2009}), respectively. 

Another reason for studying monoid controlled automata, especially in
the case of groups, is that the word problems of a group $G$ are
contained in a full trio (such as the context-free or the indexed
languages) if and only if the languages accepted by valence automata
over $G$ are contained in this full trio (see, for example,
\cite[Proposition 2]{Kambites2009}). Thus, valence automata offer an
automata theoretic interpretation of word problems for groups.

A similar situation holds
for context-free grammars where each production is assigned a monoid
element such that a derivation is valid as soon as the product of the
monoid elements (in the order of the application of the rules) is the
identity.  Here, the integers, the multiplicative group of $\Q$, and
powers of the bicyclic monoid lead to additive and multiplicative
valence grammars and Petri net controlled grammars, respectively. The
latter are in turn equivalent to matrix grammars (with erasing and
without appearance checking, see \cite{DassowTuraev09a} for details).
Therefore, the investigation of monoid control mechanisms promises
very general insights into a variety of models.

One of the most basic problems for these models is the
characterization of those monoids whose use as control mechanism
actually increases the power of the respective model.  For monoid
controlled automata, such a characterization has been achieved by
Mitrana and Stiebe \cite{MitranaStiebe2001} for the case of groups,
but has not been established for monoids.  For valence grammars, that
is, context-free grammars with monoid control, very little was known
in this respect up to date. It was an open problem whether valence
grammars over finite monoids are capable of generating languages that
are not context-free (see \cite[p. 387]{FernauStiebe2002a}). 

Another important question is for which monoids the extended automata can be
determinized, that is, for which monoids the deterministic variant is as
powerful as the nondeterministic one.  Mitrana and Stiebe \cite{MitranaStiebe2001} have shown that automata controlled by a group cannot be
determinized if the group contains at least one element of infinite order.
However, the exact class of monoids for which automata can be determinized was
not known to date.

The contribution of this work is twofold. On the one hand, the open question
of whether all languages generated by valence grammars over finite monoids are
context-free is settled affirmatively. On the other hand, we present an
algebraic dichotomy of monoids that turns out to provide a characterization for all the
conditions above. Specifically, we show that the following assertions are equivalent:
\begin{itemize}
\item Valence grammars over $M$ generate only context-free languages.
\item Valence automata over $M$ accept only regular languages.
\item Valence automata over $M$ can be determinized.
\item Valence transducers over $M$ perform only rational transductions.
\item In each finitely generated submonoid of $M$, only finitely many elements possess 
a right inverse.
\end{itemize}

\paragraph{Acknowledgements}
I would like to thank Reiner H\"{u}chting and Klaus Madlener for discussions
and helpful comments which have improved the presentation of the paper.

\section{Basic notions}
A \emph{monoid} is a set $M$ together with an associative operation and a
neutral element.  Unless defined otherwise, we will denote the neutral element
of a monoid by $1$ and its operation by juxtaposition.  That is, for a monoid
$M$ and $a,b\in M$, $ab\in M$ is their product.  
The \emph{opposite monoid} $\opp{M}$ of $M$ has the same set of elements as $M$, but has
the operation $\circ$ with $a\circ b\defeq ba$, $a,b\in M$.
For $a,b\in M$, we write
$a\sqsubseteq b$ iff there are $c,d\in M$ such that $b=ac=da$.  Let $a\in M$. An
element $b\in M$ with $ab=1$ is called a \emph{right inverse} of $a$.  If $b\in
M$ obeys $ba=1$, it is a \emph{left inverse} of $a$. An element that is both a
left and a right inverse is said to be a \emph{two-sided inverse}. By
$\One$, we denote the trivial monoid that consists of just one element.
$M$ is said to be \emph{left-cancellative} if $ab=ac$ implies $b=c$ for
$a,b,c\in M$. Whenever $\opp{M}$ is left-cancellative, we say that $M$ is
\emph{right-cancellative}.

A subset $N\subseteq M$ is said to be a \emph{submonoid of $M$} iff $1\in
N$ and $a,b\in N$ implies $ab\in N$. For a subset $N\subseteq M$, let
$\langle N\rangle$ be the intersection of all submonoids $N'$ of $M$ that
contain $N$. That is, $\langle N\rangle$ is the smallest submonoid of $M$
that contains $N$.  $\langle N\rangle$ is also called the \emph{submonoid
generated by $N$}. We call a monoid \emph{finitely generated} if it is generated by
a finite subset.
In each monoid $M$, we have the following submonoids:
\begin{eqnarray*}
\Ri{M}&\defeq \{a\in M \mid \exists b\in M: ab=1 \}, \\
\Li{M}&\defeq \{a\in M \mid \exists b\in M: ba=1 \}.
\end{eqnarray*}
The elements of $\Ri{M}$ and $\Li{M}$ are called \emph{right invertible} and
\emph{left invertible}, respectively.
In addition, for every element $a\in M$, we define the sets
\begin{eqnarray*}
\RInv{a}&\defeq \{b\in M \mid ab=1 \}, \\
\LInv{a}&\defeq \{b\in M \mid ba=1 \}. 
\end{eqnarray*}
When using a monoid $M$ as part of a control mechanism, the subset 
\[ \E{M}\defeq \{a\in M \mid \exists b,c\in M: bac=1\} \] 
will play an important role. If in $M$ every element has a two-sided
inverse, we call $M$ a \emph{group}.

Let $\Sigma$ be a fixed countable set of abstract symbols, the finite subsets
of which are called \emph{alphabets}.  For an alphabet $X$, we will write $X^*$
for the set of words over $X$.  The empty word is denoted by $\lambda\in X^*$.
In particular, $\emptyset^*=\{\lambda\}$.  Together with the concatenation as
its operation, $X^*$ is a monoid.  We will regard every $x\in X$ as an element
of $X^*$, namely the word consisting only of one occurence of $x$.  For a
symbol $x\in X$ and a word $w\in X^*$, let $|w|_x$ be the number of occurrences
of $x$ in $w$. For a subset $Y\subseteq X$, let $|w|_Y\defeq\sum_{x\in
Y}|w|_x$.  By $|w|$, we will refer to the length of $w$.  By $X^{\le
n}\subseteq X^*$, for $n\in\N$, we denote the set of all words over $X$ of
length $\le n$.  Given alphabets $X,Y$, subsets of $X^*$ and $X^*\times Y^*$
are called \emph{languages} and \emph{transductions}, respectively. We define
the \emph{shuffle} $L_1\shuf L_2$ of two languages $L_1,L_2\subseteq X^*$ to be
the set of all words $w\in X^*$ such that $w=u_1v_1\cdots u_nv_n$ for some
$u_i,v_i\in X^*$, $1\le i\le n$, with $u_1\cdots u_n\in L_1$, $v_1\cdots v_n\in
L_2$.  When $\{w\}$ is used as an operand for $\shuf$, we also just write $w$ instead of $\{w\}$.
For
$x_1,\ldots,x_n\in X$, let $\rev{(x_1\cdots x_n)}\defeq x_n\cdots x_1$.

Let $M$ be a monoid. An \emph{automaton over $M$} is a tuple $A=(M,Q,E,q_0,F)$, in which 
$Q$ is a finite set of \emph{states}, $E$ is a finite
subset of $Q\times M\times Q$, called the set of \emph{edges}, $q_0\in Q$ is the \emph{initial
state}, and $F\subseteq Q$ is the set of \emph{final states}. The \emph{step
relation} $\step{A}$ of $A$ is a binary relation on $Q\times M$, for
which $(p,a) \step{A} (q,b)$ iff there is an edge $(p,c,q)$ such that
$b=ac$. The set generated by $A$ is then
\[S(A)\defeq\{a\in M \mid \exists q\in F: (q_0,1)\step{A}^* (q,a) \}. \]
A \emph{valence automaton over $M$} is an automaton $A$ over $X^*\times
M$, where $X$ is an alphabet. 
$A$ is said to be \emph{deterministic} if all its edges are in $Q\times (X\times M)\times Q$ and, for each pair $(q,x)\in Q\times X$, there is at most one edge $(q,(x,m),p)$ for $m\in M, p\in Q$.
The \emph{language accepted by $A$} is defined as
\[ L(A)\defeq \{w\in X^* \mid (w,1)\in S(A)\}. \]
A \emph{finite automaton} is a valence automaton over $\One$.
For a finite automaton $A=(X^*\times\One, Q,E,q_0,F)$, we also write
$A=(X,Q,E,q_0,F)$.
Languages accepted by finite automata are called \emph{regular languages}.
A \emph{valence transducer over $M$} is an automaton $A$ over
$X^*\times Y^*\times M$, where $X$ and $Y$ are alphabets. The \emph{transduction
performed by $A$} is
\[T(A)\defeq \{(x,y)\in X^*\times Y^* \mid (x,y,1)\in S(A) \}.\]
A \emph{finite state transducer} is a valence transducer over
$\One$. 
For a finite state transducer $A=(X^*\times Y^*\times\One, Q,E,q_0,F)$, we also write
$A=(X,Y,Q,E,q_0,F)$.
Transductions performed by finite state transducers are called
\emph{rational transductions}.

A \emph{valence grammar over $M$} is a tuple $G=(N,T,M,P,S)$, where $N,T$ are
disjoint alphabets, called the \emph{nonterminal} and \emph{terminal
alphabet}, respectively, $P\subseteq N\times (N\cup T)^*\times M$ is a finite set of
\emph{productions}, and $S\in N$ is the \emph{start symbol}. 
For a production $(A,w,m)\in P$ we also write $(A\to w; m)$.
The
\emph{derivation relation} $\step{G}$ of $G$ is a binary relation on $(N\cup
T)^*\times M$, for which $(u,a)\step{G}(v,b)$ iff there is a $(A\to w;c)\in P$
and words $r,s\in (N\cup T)^*$ such that $u=rAs$, $v=rws$, and $b=ac$. 
The \emph{language generated by $G$} is defined as
\[ L(G)\defeq \{w\in T^* \mid (S,1) \step{G}^* (w,1) \}. \]
Valence grammars were introduced by P\u{a}un in \cite{Paun1980}. A thorough
treatment, including normal form results and a classification of the resulting language classes for commutative
monoids, has been carried out by Fernau and Stiebe \cite{FernauStiebe2002a}.
Valence grammars over $\One$ are called \emph{context-free grammars}. For a
context-free grammar $G=(N,T,\One,P,S)$, we also write $G=(N,T,P,S)$.
Furthermore, a production $(A\to w; 1)\in P$ in a context-free grammar is also
written $A\to w$.  Languages generated by context-free grammars are called
\emph{context-free}.

\section{A dichotomy of monoids}

An \emph{infinite ascending chain} in $M$ is an infinite sequence $x_1,x_2,\ldots$ of pairwise distinct elements of $M$
such that $x_i\sqsubseteq x_{i+1}$ for all $i\in\N$.
\begin{lemma}\label{lemma:cdichotomy}
Let $M$ be left- or right-cancellative. Then exactly one of the following holds:
\begin{enumerate}
\item $M$ is a finite group.
\item $M$ contains an infinite ascending chain.
\end{enumerate}
\end{lemma}
\begin{proof}
Suppose $M$ does not contain an infinite ascending chain.

First, we prove that $M$ is a group. Let $r\in M$ and consider the elements
$r^i\in M$, $i\in\N$. Since $r^i\sqsubseteq r^j$ for $i\le j$, our assumption implies
that there are $i,j\in\N$, $i<j$, with $r^i=r^j$. Since $M$ is left- or
right-cancellative, this implies $r^{j-i}=1$, meaning that $r$ has in
$r^{j-i-1}$ a two-sided inverse. Thus, $M$ is a group.

This implies that $r\sqsubseteq s$ for any $r,s\in M$. Therefore, if
$M$ were infinite, it would contain an infinite ascending
chain. 
\end{proof}

\begin{lemma}\label{lemma:disjoint} 
Let $s,t\in M$, $s\ne t$, and $s\sqsubseteq t$. Then,
$\RInv{s}\cap\RInv{t}=\emptyset$ and $\LInv{s}\cap\LInv{t}=\emptyset$.
\end{lemma} 
\begin{proof}
We only show $\RInv{s}\cap\RInv{t}=\emptyset$ since then
$\LInv{s}\cap\LInv{t}=\emptyset$ follows by applying the former to the opposite
monoid.  Write $t=us$, $u\in M$, and suppose there were a
$z\in\RInv{s}\cap\RInv{t}$.  Then, $1=tz=usz=u$ and thus $t=s$. 
\end{proof}

\begin{theorem}\label{thm:dichotomy}
For every monoid $M$, exactly one of the following holds:
\begin{enumerate}
\item The subsets $\Ri{M}$, $\Li{M}$, and $\E{M}$ coincide and constitute a
finite group.
\item $\Ri{M}$ and $\Li{M}$ each contain an infinite ascending chain. In particular, there exist
infinite sets $S\subseteq\Ri{M}$ and $S'\subseteq\Li{M}$ such that
$\RInv{s}\cap\RInv{t}=\emptyset$ for $s,t\in S$, $s\ne t$, and
$\LInv{s'}\cap\LInv{t'}=\emptyset$ for $s',t'\in S'$, $s'\ne t'$.
\end{enumerate}
\end{theorem}
\begin{proof}
First, we claim that $\Ri{M}$ is infinite if and only if $\Li{M}$ is infinite.
Here, it suffices that $\Ri{M}$ being infinite implies the infinity of
$\Li{M}$, since the other direction follows by considering the opposite monoid.
If $\Ri{M}$ is infinite, it contains an infinite ascending chain according to
Lemma \ref{lemma:cdichotomy}. By Lemma \ref{lemma:disjoint}, the elements of
the chain have pairwise disjoint sets of right inverses, which are non-empty.
Since right inverses are left invertible, $\Li{M}$ is infinite.

Suppose $\Ri{M}$ and $\Li{M}$ are both finite.  Since $\Ri{M}$ is
right-cancellative, it is a group by Lemma \ref{lemma:cdichotomy}.  Thus, we
have $\Ri{M}\subseteq\Li{M}$ and analogously $\Li{M}\subseteq\Ri{M}$. In order
to prove $\E{M}=\Ri{M}$, we observe that $\Ri{M}\subseteq\E{M}$ by definition.
Now, suppose $a\in\E{M}$ to be witnessed by $bac=1$, $b,c\in M$.  By this
equation, we have $b\in\Ri{M}$ and can multiply $b^{-1}$ on the left and then
$b$ on the right. We obtain $acb=1$ and thus $a\in\Ri{M}$. This proves that
$\Ri{M}=\Li{M}=\E{M}$ and that this is a finite group.

In case $\Ri{M}$ and $\Li{M}$ are both infinite, the infinite ascending chains
are provided by Lemma \ref{lemma:cdichotomy}.  By Lemma \ref{lemma:disjoint},
their elements form sets $S\subseteq\Ri{M}$ and $S'\subseteq\Li{M}$ with the
desired properties.
\end{proof}

\section{Capabilities of valence automata and transducers}
In this section, we show that the following conditions are equivalent:
\begin{itemize}
\item Valence automata over $M$ accept only regular languages.
\item Valence automata over $M$ can be determinized.
\item Valence transducers over $M$ perform only rational transductions.
\item $\Ri{N}$ is finite for every finitely generated submonoid $N$ of $M$. 
\end{itemize}

\begin{lemma}\label{lemma:rfintoauto}
Let $\Ri{N}$ be finite for every finitely generated submonoid $N$ of $M$.  Then,
valence automata over $M$ accept only regular languages and valence transducers
over $M$ perform only rational transductions. In particular, valence automata
over $M$ can be determinized.
\end{lemma}
\begin{proof}
Let $A=(X^*\times M, Q,E,q_0,F)$ be a valence automaton over $M$. Since $E$
is finite, the set of $m\in M$ such that there is some edge $(p,(w,m),q)$
in $E$ is finite. If $N$ is the submonoid of $M$ generated by these $m\in
M$, we can regard $A$ as a valence automaton over $N$. Thus, let
$A=(X^*\times N, Q, E, q_0, F)$. Furthermore, removing edges of the form
$(p,(w,m),q)$ such that $m\notin \E{N}$ will not alter the accepted
language, since such edges cannot be used in a successful run.  By Theorem
\ref{thm:dichotomy}, $\E{N}$ is a finite group and we can
assume $A=(X^*\times \E{N}, Q, E, q_0, F)$. Since
$\E{N}$ is finite, a finite automaton accepting $L(A)$ can now easily be
constructed by incorporating the monoid elements into the states.
The proof for the valence transducers over $M$ works completely analogously.

Since finite automata can be determinized and we have seen that valence
automata over $M$ accept only regular languages, it follows that valence
automata over $M$ can be determinized.
\end{proof}

In \cite{MitranaStiebe2001}, Mitrana and Stiebe proved that valence automata
over groups with at least one element of infinite order cannot be determinized.
We can now use a similar idea and our dichotomy theorem to provide a
characterization of those monoids over which valence automata can be
determinized. 
\begin{lemma}\label{lemma:autotorfin}
Let $M$ be a finitely generated monoid such that $\Ri{M}$ is infinite. Then,
there is a valence automaton over $M$ whose accepted language cannot be
accepted by a deterministic valence automaton over $M$. In particular, valence
automata over $M$ can accept non-regular languages and valence transducers over
$M$ can perform non-rational transductions.
\end{lemma}
\begin{proof}
Let $M$ be generated by the finite set $\{a_1,\ldots,a_n\}$ and let $X=\{ x_1,\ldots,x_n\}$,
$Y=\{y_1,\ldots,y_n\}$ be disjoint alphabets. Let $\varphi:(X\cup Y)^*\to M$ be
the epimorphism defined by $\varphi(x_i)\defeq\varphi(y_i)\defeq a_i$ and
$K\defeq X^*\cup \{w\in X^*Y^* \mid \varphi(w)=1\}$. Then, $K$ is 
clearly accepted by a (nondeterministic) valence automaton over $M$.  Suppose $K$
were accepted by a deterministic valence automaton $A$ over $M$. Let
$S\subseteq\Ri{M}$ be the infinite set provided by Theorem \ref{thm:dichotomy}.
The infinity of $S$ implies that we can find an infinite set $S'\subseteq X^*$
such that $\varphi(S')=S$ and $\varphi(u)\ne\varphi(v)$ for $u,v\in S'$, $u\ne
v$.  Since $A$ is deterministic and $S'\subseteq L(A)$, each word $w\in S'$
causes $A$ to enter a configuration $(q(w), 1)$, where $q(w)$ is a final state.
Choose $u,v\in S'$ such that $u\ne v$ and $q(u)=q(v)$.  Let $u'\in Y^*$ be a word such that
$\varphi(u)\varphi(u')=1$, this is possible since $\varphi(u)\in\Ri{M}$ and
$\varphi$ is surjective. The word $u'$ causes $A$ to go from
$(q(u),1)=(q(v),1)$ to $(q,1)$ for some final state $q$, since $uu'\in K$.
Thus, $vu'$ is also contained in $K$ and hence $\varphi(v)\varphi(u')=1$, but
$\RInv{\varphi(u)}\cap\RInv{\varphi(v)}=\emptyset$, a contradiction.

Thus, $K$ is not accepted by a deterministic valence automaton over $M$. In
particular, $K$ is not regular.  Furthermore, from the valence automaton accepting $K$,
a valence transducer can be constructed that maps $\{\lambda\}$ to $K$. Since
$K$ is not regular, the transduction performed by the transducer is not
rational.
\end{proof}

\section{Capabilities of valence grammars}
In this section, it is shown that the following conditions are equivalent:
\begin{enumerate}
\item Valence grammars over $M$ generate only context-free languages.
\item $\Ri{N}$ is finite for every finitely generated submonoid $N$ of $M$.
\end{enumerate}
In one of the directions, we have to construct a context-free grammar for
valence grammars over monoids that fulfill the second condition.  Because of
the limited means available in the context-free case, the constructed grammar
can simulate only a certain fragment of the derivations in the valence grammar.
Thus, we will have to make sure that every word generated by the valence
grammar has a derivation in the aforementioned fragment.  These derivations are
obtained by considering the derivation tree of a given derivation and then
choosing a suitable linear extension of the tree order. The 
construction of these linear extensions can already be described for a simpler kind of
partial order, \emph{valence trees}.

\newcommand{\depth}{excursiveness}
Let $X$ be an alphabet and $U\subseteq X$ a subset. Then, each word $w\in X^*$
has a unique decomposition $w=y_0x_1y_1\cdots x_ny_n$ such that $y_0,y_n\in
(X\setminus U)^*$, $y_i\in (X\setminus U)^+$ for $1\le i\le n-1$ and $x_i\in
U^+$ for $1\le i\le n$.  This decomposition is called \emph{$U$-decomposition}
of $w$ and we define $\factors{U}{w}\defeq n$. 

A \emph{tree} is a finite partially ordered set $(\T,\le)$ that has a least
element and where, for each $t\in \T$, the set $\{t'\in \T\mid t'\le t\}$ is
totally ordered by $\le$.  The least element is also called the \emph{root} and the maximal elements are called \emph{leaves}. A
\emph{valence tree} $\T$ over $M$ is a tuple $(\T,\le,\varphi)$, where
$(\T,\le)$ is a tree and $\varphi:\T^*\to M$ is a homomorphism\footnote{We will often assume, without loss of generality,
that $\T$ is an alphabet.} assigning a \emph{valence} to each
node.  An \emph{evaluation} defines an order in which the nodes in a valence
tree can be traversed that is compatible with the tree order.  Thus, an
\emph{evaluation} of $\T$ is a linear extension $\preceq$ of $(\T,\le)$.  Let
$w\in \T^*$ correspond to $\preceq$, i.e., let $\T=\{t_1,\ldots,t_n\}$ such
that $t_1\preceq\cdots\preceq t_n$ and $w=t_1\cdots t_n$. Then the \emph{value} of $\preceq$ is defined to be $\varphi(w)$.  An element
$v\in M$ is called a \emph{value of $\T$} if there exists an evaluation of
$(\T,\le)$ with value $v$.  Given a node $t\in \T$, let $U_t\defeq\{t'\in
\T\mid t\le t'\}$.  If $w=y_0x_1y_1\cdots x_ny_n$ is the $U_t$-decomposition of
$w$, then $\varphi(x_1),\ldots,\varphi(x_n)$ is called the \emph{valence
sequence} of $t$ in $w$ and $n$ its \emph{length}.  By the \emph{\depth{}} of
an evaluation, we refer to the maximal length of a valence sequence.  Hence,
the \depth{} of an evaluation is the maximal number of times one has to enter
any given subtree when traversing the nodes in the order given by the
evaluation. 
We are interested in finding evaluations of valence trees with small \depth{}.
Of course, for every valence tree, there are evaluations with \depth{} one
(take, for example, the order induced by a preorder traversal), but these might
not be able to cover all possible values. However, we will see in Lemma
\ref{lemma:ssequences} that, in the case of a finite group, there exists a bound
$m$ such that every value can be attained by an evaluation of \depth{} at most
$m$.

\begin{lemma}\label{lemma:commute}
For each finite group $G$, there is a constant $m\in\N$ with the following
property: For elements $g_i,h_i\in G$, $i=1,\ldots,n$, $n\ge m$,
there are indices $k,\ell\in \N$, $1\le k<\ell\le n$, such that
$g_kh_k\cdots g_{\ell}h_{\ell}=g_k\cdots g_{\ell}h_k\cdots h_{\ell}$.
\end{lemma}

\begin{proof}
Let $m=2(|G|^3+1)$ and $D\subseteq\{1,\ldots,n\}$ be the set of odd indices. 
Define the map $\alpha:D\to G^3$ by
$\alpha(i) \defeq (g_1\cdots g_i,~h_1\cdots h_i,~g_1h_1\cdots g_ih_i)$
for $i\in D$.
Since $|D|\ge |G|^3+1$, there are indices $i,j\in D$, $i<j$, such that
$\alpha(i)=\alpha(j)$. This means that
$g_{i+1}\cdots g_j=1$,~ $h_{i+1}\cdots h_j=1$, and $g_{i+1}h_{i+1}\cdots g_jh_j=1$.
Since $i,j$ are both odd, letting $k=i+1$ and $\ell=j$ implies $k<\ell$ and
yields the desired equality.
\end{proof}

\begin{lemma}\label{lemma:words}
Let $X$ be an alphabet and $U,V\subseteq X$ subsets such that either
$U\subseteq V$, $V\subseteq U$, or $U\cap V=\emptyset$.  Furthermore, let 
$r\in X^*U$, $x\in U^+$, $y\in (X\setminus U)^+$, and $s\in X^*\setminus UX^*$.
Then, we have $\factors{V}{rxys}\le \factors{V}{ryxs}$.
\end{lemma}
\begin{proof}
Suppose $V\subseteq U$. Since $y$ does not contain any symbols in $V$, we have
\begin{align*}
\factors{V}{ryxs}&=\factors{V}{r}+\factors{V}{x}+\factors{V}{s}, \\
\factors{V}{rxys}&=\factors{V}{rx}+\factors{V}{s}.
\end{align*}
Thus,
\begin{align*}
\factors{V}{rxys}&=\factors{V}{rx}+\factors{V}{s} \\
                 &\le \factors{V}{r}+\factors{V}{x}+\factors{V}{s} \\
                 &=\factors{V}{ryxs}.
\end{align*}
In the case $U\cap V=\emptyset$, $x$ does not contain any symbol in $V$. Hence,
\begin{align*}
\factors{V}{ryxs}&=\factors{V}{r}+\factors{V}{y}+\factors{V}{s}, \\
\factors{V}{rxys}&=\factors{V}{r}+\factors{V}{ys},
\end{align*}
which implies
\begin{align*}
\factors{V}{rxys}&=\factors{V}{r}+\factors{V}{ys} \\
                 &\le \factors{V}{r}+\factors{V}{y}+\factors{V}{s} \\
                 &=\factors{V}{ryxs}.
\end{align*}
Now suppose $U\subseteq V$. Since the rightmost letter of $r$ is in $V$ and $x$ lies in $V^+$, we have $\factors{V}{rxys}=\factors{V}{rys}$. Thus,
$\factors{V}{rxys}=\factors{V}{rys}\le\factors{V}{ryxs}$.
\end{proof}

\begin{lemma}\label{lemma:ssequences}
For each finite group $G$, there is a constant $m$ such that each value of a
valence tree over $G$ has an evaluation of \depth{} at most $m$.
\end{lemma}
\begin{proof}
For an alphabet $X$, we denote the set of \emph{multisets} over $X$, i.e., maps
$X\to\N$, by $X^\oplus$.  $X^\oplus$ carries a (commutative) monoid structure
by way of $(\alpha+\beta)(x)\defeq \alpha(x)+\beta(x)$ for $x\in X$.
To every evaluation $w$ of $(\T,\le)$, we assign the multiset $\mu_w\in
\T^\oplus$, which is defined by $\mu_w(t)\defeq\factors{U_t}{w}$ for every $t\in \T$.
That is, $\mu_w(t)$ is the length of the valence sequence of $t$ in $w$.

Let $m$ be the constant provided by Lemma \ref{lemma:commute} and let
$w\in\T^*$ be an
evaluation of $(\T,\le)$ such that $\mu_w$ is minimal with respect to
$\sqsubseteq$ among all evaluations with value $v$.  If we can prove that
$\mu_w(t)\le m$ for all $t\in \T$, the lemma follows.  Therefore, suppose 
that there is a $t\in \T$ with $n\defeq\mu_w(t)>m$.
Specifically, let $w=y_0x_1y_1\cdots x_ny_n$
be the $U_t$-decomposition of $w$. 
Use Lemma \ref{lemma:commute} to find indices $1\le k<\ell\le n$ with
\begin{equation}
\varphi(x_k)\varphi(y_k)\cdots \varphi(x_\ell)\varphi(y_\ell)=\varphi(x_k)\cdots \varphi(x_\ell) \varphi(y_k)\cdots \varphi(y_\ell). \label{eq:commute}
\end{equation}
Furthermore, let 
\begin{equation} w'\defeq (y_0x_1y_1\cdots x_{k-1}y_{k-1})(x_k\cdots x_{\ell} y_k\cdots y_{\ell})(x_{\ell+1}y_{\ell+1}\cdots x_ny_n). \label{eq:neweval}\end{equation}
That is, we obtain $w'$ from $w$ by replacing $x_ky_k\cdots x_\ell y_\ell$ with $x_k\cdots x_\ell y_k\ldots
y_\ell$. Then, \eqref{eq:commute} means that $\varphi(w')=\varphi(w)$.
We shall prove that $w'$ is an evaluation of $(\T,\le)$ and obeys
$\mu_{w'}\sqsubset\mu_w$, which contradicts the choice of $w$.

First, we prove that $w'$ is an evaluation. Let $u_1,u_2\in \T$ be nodes with
$u_1\le u_2$.  If $u_1<t$, then $u_1$ appears in $y_0$, and thus $u_2$ is on
the right side of $u_1$ in $w'$.  If $u_1\ge t$, then each of the nodes
$u_1,u_2$ appears in some $x_i$ and therefore do not change their relative
positions.  If $u_1$ and $t$ are incomparable, then $u_2$ and $t$ are also
incomparable and each of $u_1,u_2$ appears in some $y_i$.  Again, $u_1$ and
$u_2$ do not change their relative positions. Thus, $w'$ corresponds to a linear
extension of $\le$.

We want to show that $\mu_{w'}\sqsubseteq \mu_w$.  To this end, we consider the
words
\[ w_i\defeq (y_0x_1y_1\cdots x_{k-1}y_{k-1})(x_k\cdots x_{k+i}y_k\cdots y_{k+i})(x_{k+i+1}y_{k+i+1}\cdots x_ny_n) \]
for $0\le i\le \ell-k$. With these, we have $w=w_0$ and $w'=w_{\ell-k}$. Since $(\T,\le)$ is a tree, we have
$U_u\subseteq U_t$, $U_t\subseteq U_u$, or $U_u\cap U_t=\emptyset$ for every
$u\in \T$. Therefore, we can apply Lemma \ref{lemma:words} 
to $U\defeq U_t$, $V\defeq U_u$, and
\begin{align*}
r&\defeq (y_0x_1y_1\cdots x_{k-1}y_{k-1})(x_k\cdots x_{k+i}),	& x&\defeq x_{k+i+1}, \\
y&\defeq y_k\cdots y_{k+i},              			& s&\defeq y_{k+i+1}(x_{k+i+2}y_{k+i+2}\cdots x_ny_n),
\end{align*}
which yields $\factors{U_u}{w_{i+1}}\le \factors{U_u}{w_i}$ for $0\le i<\ell-k$. This implies 
$\mu_{w'}(u)\le \mu_w(u)$ and therefore $\mu_{w'}\sqsubseteq\mu_w$.

It remains to be shown that $\mu_{w'}$ is strictly smaller than $\mu_w$.
In $w'$, the node $t$ has the valence sequence
\[\varphi(x_1),\ldots, \varphi(x_{k-1}),\varphi(x_k\cdots x_\ell),\varphi(x_{\ell+1}),\cdots \varphi(x_n),\]
which has length $\mu_{w'}(t)=n-(\ell-k)<n=\mu_w(t)$. 
\end{proof}

\newcommand{\valencemap}{\varphi}
\newcommand{\labelmap}{\Lambda}

We define a \emph{derivation tree} for a valence grammar $G=(N,T,M,P,S)$ to be
a tuple $(\T,\le,\valencemap,(\le_t)_{t\in\T},\labelmap)$, where
\begin{itemize}
\item $(\T,\le,\valencemap)$ is a valence tree,
\item for each $t\in\T$, $\le_t$ is a total order on the set of successors of $t$,
\item $\labelmap:\T\to N\cup T\cup \{\lambda\}$ defines a \emph{label} for each node,
\item if $t\in\T$ is a node with the successors $s_1,\ldots,s_n$ such that
$s_1\le_t \ldots \le_t s_n$, then we either have $\labelmap(t)\in T\cup
\{\lambda\}$, $n=0$, and $\varphi(t)=1$ or we have $\labelmap(t)\in N$ and
there is a production
$(\labelmap(t)\to\labelmap(s_1)\cdots\labelmap(s_n);\valencemap(t))$ in $P$.
\end{itemize}
The total orders $\le_t$, $t\in\T$, induce a total order on the set of leaves
(see \cite[Section 4.3]{HopcroftUllman1979} for details), which in turn
defines a word $w\in T^*$. This word is called the \emph{yield} of the
derivation tree.

Each derivation tree can be regarded as a valence tree.
An evaluation then defines a derivation $(A,1)\step{G}^*(w,v)$, where $A\in N$
is the label of the root, $w$ is the yield, and $v\in M$ is the value of the
evaluation.  Conversely, every derivation induces a derivation tree and an
evaluation.  Thus, a word $w\in T^*$ is in $L(G)$ iff there exists a
derivation tree for $G$ with yield $w$, a root labeled $S$, and an evaluation
with value $1$.  See \cite[Section 4.2]{FernauStiebe2002a} for details.

\newcommand{\concat}{\Box}
\begin{lemma}\label{lemma:rfintogrammar}
Let $\Ri{N}$ be finite for every finitely generated submonoid $N$ of $M$.
Furthermore, let $G=(N,T,M,P,S)$ be a valence grammar over $M$. Then, $L(G)$ is
context-free.
\end{lemma}
\begin{proof}
As in the proof of Lemma \ref{lemma:rfintoauto}, we can assume that $M$ is finitely
generated and thus has a finite $\Ri{M}$.  Since productions $(A\to w;m)$ with
$m\notin\E{M}$ cannot be part of a successful derivation, their removal does
not change the generated language.  Furthermore, by Theorem
\ref{thm:dichotomy}, $\E{M}$ is a finite group. Thus, we can assume that
$G=(N,T,H,P,S)$, where $H=\E{M}$ is a finite group.  By a simple construction,
we can further assume that in $G$, every production is of the form $(A\to w;
h)$ with $w\in N^*$ or $(A\to w; 1)$ with $w\in T\cup\{\lambda\}$. 

We shall construct a context-free grammar $G'=(N',T,P',S')$ for $L(G)$. The
basic idea is that $G'$ will simulate derivations of bounded \depth{}. This is
done by letting the nonterminals in $G'$ consist of a nonterminal
$A\in N$
and a finite sequence $\sigma$ of elements from $H$.  $G'$ then simulates the
generation of a nonterminal $A$ by generating a pair $(A,\sigma)$ and thereby
guesses that the corresponding node in the derivation tree of $G$ will have
$\sigma$ as its valence sequence.  Lemma \ref{lemma:ssequences} will then
guarantee that this allows $G'$ to derive all words in $L(G)$ when the
sequences $\sigma$ are of bounded length.

\newcommand{\join}{J}
Formally, we will regard $H$ as an alphabet and a sequence will be a word over $H$. 
In order to be able to distinguish between the concatenation of words in $H^*$
and the group operation in $H$, we will denote the concatenation in $H^*$ by
$\concat$. Thus, let $N'=N\times H^{\le m}$, in which $m\in\N$ is the
constant provided by Lemma \ref{lemma:ssequences} for the group $H$.
The set of sequences that can be obtained from another sequence $\sigma$ by
``joining'' subsequences is denoted by
$\join(\sigma)$:
\[ \join(h_1\concat h_2\concat \sigma)\defeq \join((h_1h_2)\concat\sigma)\cup \{h_1\concat \sigma' \mid \sigma'\in \join(h_2\concat\sigma) \} \]
for $h_1,h_2\in H$ and $\sigma\in H^*$ and $\join(\sigma)\defeq\{\sigma\}$ if $|\sigma|\le 1$.
$J$ is defined for subsets $S\subseteq H^*$ by $J(S)\defeq \bigcup_{\sigma\in S} J(\sigma)$.

For each production $(A\to w;h)\in P$, $w=B_1\cdots B_n$, $B_i\in N$ for $1\le i\le n$, we include the production 
\[(A,\sigma)\to(B_1,\sigma_1)\cdots(B_n,\sigma_n),\]
for each $\sigma\in H^{\le m}\setminus \{\lambda\}$ and $\sigma_1,\ldots,\sigma_n\in H^{\le m}$ such that
for $\sigma=h_1\concat\sigma'$, $h_1\in H$, $\sigma'\in H^{\le m-1}$, one of the following
holds:
\begin{itemize}
\item $(h^{-1}h_1)\concat \sigma'\in\join(\sigma_1\shuf \cdots \shuf \sigma_n)$.
\item $h_1=h$ and $\sigma'\in\join(\sigma_1\shuf \cdots \shuf \sigma_n)$.
\end{itemize}
Furthermore, for every production $(A\to w, 1)$, $w\in T\cup\{\lambda\}$, we include $(A,\lambda)\to w$.
Finally, the start symbol of $G'$ is $(S,1)$.

It remains to be shown that $L(G')=L(G)$. In order to prove $L(G')\subseteq
L(G)$, one can show by induction on $n$ that for $w\in T^*$, $(A,\sigma)\step{G'}^n w$ implies
that there is a derivation $(A,1)\step{G}^* (w,h)$ for some $h\in H$ using productions $(A_1\to
w_1;h_1),\ldots,(A_k\to w_k;h_k)$ such that $\sigma\in J(h_1\concat
\cdots\concat h_k)$. This implies that for $(S,1)\step{G'}^* w$, $w\in
T^*$, we have $w\in L(G)$. Thus, $L(G')\subseteq L(G)$.

Let $w\in L(G)$ with derivation tree
$(\T,\le,\varphi,(\le_t)_{t\in\T},\Lambda)$. By Lemma \ref{lemma:ssequences},
there is an evaluation $\preceq$ of the tree of \depth{} $\le m$. From the tree
and the evaluation, we construct a derivation tree
$(\T,\le,\varphi',(\le_t)_{t\in\T},\Lambda')$ for $w$ in $G'$ as follows. The
components $\T$, $\le$, and $\le_t$, $t\in\T$, stay unaltered, but $\varphi'$
will assign $1$ to each node and $\Lambda'$ is defined by
$\Lambda'(t)\defeq\Lambda(t)$ if $\Lambda(t)\in T\cup \{\lambda\}$ and 
$\Lambda'(t) \defeq (\Lambda(t), h_1\concat \cdots \concat h_k)$ if $\Lambda(t)\in
N$, where $h_1,\ldots,h_k$ is the valence sequence of $t$ in $\preceq$.
Now, one can see that the new tree is a derivation tree for $G'$ that
generates $w$ with any evaluation. Hence, $L(G)\subseteq L(G')$.
\end{proof}

In order to prove the main result of this section, we need to exhibit a
valence grammar over $M$ that generates a non-context-free language when
given a finitely generated monoid $M$ with infinite $\Ri{M}$. In the proof
that the generated language is not context-free, we will use the following
well-known Iteration Lemma by Ogden \cite{Ogden1968}.
\begin{lemma}[Ogden]\label{lemma:ogden}
For each context-free language $L$, there is an integer $m$ such that for
any word $z\in L$ and any choice of at least $m$ distinct marked positions
in $z$, there is a decomposition $z=uvwxy$ such that:
\begin{enumerate}
\item\label{ogden:wgeone} $w$ contains at least one marked position.
\item\label{ogden:either} Either $u$ and $v$ both contain marked positions, or
$x$ and $y$ both contain marked positions.
\item\label{ogden:most} $vwx$ contains at most $m$ marked positions.
\item\label{ogden:pump} $uv^iwx^iy\in L$ for every $i\ge 0$.
\end{enumerate}
\end{lemma}

\begin{lemma}\label{lemma:grammartorfin}
Let $\Ri{M}$ be infinite for some finitely generated monoid $M$.  Then,
there is a valence grammar over $M$ that generates a language that is not
context-free.
\end{lemma}
\begin{proof}
Let $M$ be generated by $a_1,\ldots,a_n$ and let $X=\{x_1,\ldots,x_n\}$ be
an alphabet. Furthermore, let $\varphi:X^*\to M$ be the surjective
homomorphism defined by $\varphi(x_i)=a_i$. The valence grammar
$G=(N,T,M,P,S_0)$ is defined as follows. Let $N=\{S_0, S_1\}$,
$T=X\cup\{c\}$, and let $P$ consist of the productions
\[ (S_0\to x_iS_0x_i, a_i),~~(S_0\to cS_1c, 1),~~(S_1\to x_iS_1, a_i), ~~(S_1\to\lambda,1) \]
for $1\le i\le n$. Then, clearly $L(G)=K\defeq\{rcsc\rev{r} \mid
r,s\in X^*,~\varphi(rs)=1 \}$.
It remains to be shown that $K$ is not context-free.
Suppose $K$ is context-free and let $m$ be the constant provided by Lemma 
\ref{lemma:ogden}.
By Theorem \ref{thm:dichotomy}, we can find an infinite subset
$S\subseteq\Li{M}$ such that $\LInv{a}\cap\LInv{b}=\emptyset$ for $a,b\in S$,
$a\ne b$. Since $\varphi$ is surjective, we can define $\ell(a)$ for every
$a\in S$ to be the minimal length of a word $w\in X^*$ such that
$\varphi(w)a=1$.  If $\ell(a)<m$ for all $a\in S$, the finite set $\{\varphi(w)
\mid w\in X^*, |w|<m\}$ contains a left inverse for every $a\in S$. This,
however, contradicts the fact that the infinitely many elements of $S$ have
disjoint sets of left inverses. Thus, there exists an $a\in S$ with $\ell(a)\ge
m$. We choose words $r,s\in X^*$ such that $\varphi(s)=a$ and $r$ is of minimal
length among those words satisfying $\varphi(rs)=1$.  Then, by the choice of
$a$, we have $|r|\ge m$.  

We apply the Iteration Lemma to the word
$z=rcsc\rev{r}\in K$, where we choose the first $|r|$ symbols to be marked. Let
$z=uvwxy$ be the decomposition from the lemma.  Condition \ref{ogden:wgeone}
implies $|uv|<|r|$. Because of \ref{ogden:pump}, $x$ cannot contain a $c$.
Furthermore, $x$ cannot be a subword of $r$, since then pumping would
lead to words with mismatching first and third segment. In particular, from
condition \ref{ogden:either}, the first part holds and $v$ is not empty.  Thus,
if $x$ were a subword of $s$, pumping would again lead to a mismatching first
and third segment. Hence, $x$ is a subword of $\rev{r}$.  If we now pump with
$i=0$, we obtain a word $r'cscr''\in K$, where $|r'|<|r|$. In particular, we
have $\varphi(r's)=1$, in contradiction to the choice of $r$. 
\end{proof}

\begin{theorem}\label{thm:equivalence}
Let $M$ be a monoid. The following conditions are equivalent:
\begin{enumerate}
\item\label{ass:grammars} Valence grammars over $M$ generate only context-free languages.
\item\label{ass:automata} Valence automata over $M$ accept only regular languages.
\item\label{ass:det} Valence automata over $M$ can be determinized.
\item\label{ass:transducers} Valence transducers over $M$ perform only rational transductions.
\item\label{ass:rfin} $\Ri{N}$ is finite for every finitely generated submonoid $N$ of $M$.
\item\label{ass:lfin} $\Li{N}$ is finite for every finitely generated submonoid $N$ of $M$.
\item\label{ass:efin} $\E{N}$ is finite for every finitely generated submonoid $N$ of $M$.
\end{enumerate}
\end{theorem}
\begin{proof}
Theorem \ref{thm:dichotomy} immediately implies that
\ref{ass:rfin}, \ref{ass:lfin} and \ref{ass:efin} are equivalent. 
\ref{ass:grammars} is equivalent to \ref{ass:rfin} by Lemma
\ref{lemma:grammartorfin} and Lemma \ref{lemma:rfintogrammar}.
Lemmas \ref{lemma:autotorfin} and \ref{lemma:rfintoauto} prove that
\ref{ass:automata}, \ref{ass:det}, and \ref{ass:transducers} are each equivalent to \ref{ass:rfin}. 
\end{proof}

\bibliographystyle{alpha}
\bibliography{bibliography.bib}

\end{document}